\documentclass[%
 reprint,
superscriptaddress,
nofootinbib,
amsmath,amssymb, amsfonts,
showkeys,
prx,
floatfix
]{revtex4-2}

\usepackage{graphicx}
\usepackage{dcolumn}
\usepackage{bm}
\usepackage{amsmath}
\usepackage[dvipsnames]{xcolor} 

\usepackage{float}
\usepackage{placeins}
\usepackage{silence}
\usepackage[
    colorlinks,
    citecolor = RoyalBlue,
    linkcolor = RoyalBlue,
    urlcolor = RoyalBlue,
]{hyperref}

\newcommand\thefontsize{The current font size is: \f@size pt}
\newcommand{\ket}[1]{\lvert #1 \rangle}
\newcommand{\mb}[1]{\mathbf{#1}}
\newcommand{\te}[1]{\hat{\bm{#1}}}
\newcommand{\ste}[1]{\hat{\mathbb{#1}}}
\newcommand{\eps}{\varepsilon}
\newcommand{\om}{\omega}
\newcommand{\Om}{\Omega}

\newcommand{\Eq}[1]{Eq.~(\ref{#1})}
\begin{document}
\raggedbottom 

\title{Resonant states of structured photonic time crystals}

\author{Adri\`a Can\'os Valero}
\email{adria.canos-valero@uni-graz.at}
\affiliation{Institute of Physics, University of Graz, and NAWI Graz, 8010 Graz, Austria }
\affiliation{Riga Technical University, Institute of Telecommunications, Riga, 1048, Latvia}

\author{Sergei Gladyshev}
\affiliation{Institute of Physics, University of Graz, and NAWI Graz, 8010 Graz, Austria }

\author{David Globosits}
\affiliation{Institute for Theoretical Physics, Vienna University of Technology (TU Wien), 1040 Vienna, Austria}

\author{Stefan Rotter}
\affiliation{Institute for Theoretical Physics, Vienna University of Technology (TU Wien), 1040 Vienna, Austria}

\author{Egor A. Muljarov}
\affiliation{School of Physics and Astronomy, Cardiff University, Cardiff CF24 3AA, United Kingdom}

\author{Thomas Weiss}
\email{thomas.weiss@uni-graz.at}
\affiliation{Institute of Physics, University of Graz, and NAWI Graz, 8010 Graz, Austria }

\begin{abstract}

    
    Photonic time crystals (PTCs) are spatially uniform media with periodic modulation in time, enabling momentum bandgaps and the parametric amplification of light. While their potential in optical systems is very promising, practical implementations require temporally modulating complex nanostructures of finite size, for which the physics is no longer governed by bulk properties but by resonant states, or quasinormal modes. Despite their importance, a quantitative theory describing the dynamics of these modes has been missing–a gap we address here by developing a comprehensive resonant state theory for PTCs with arbitrary geometry. Our framework provides a detailed understanding of the resonant behavior of ``structured'' PTCs and uncovers several fundamental phenomena. For weak modulations, we find a universal quadratic dependence of the eigenfrequencies on the modulation amplitude. Moreover, each static resonant mode gives rise to an infinite ladder of new eigenmodes, spaced by integer multiples of the modulation frequency. Crucially, we show that parametric amplification in these systems arises from a fundamentally resonant process, not captured by the momentum bandgap picture of ``bulk'' PTCs. We apply our theory to a realistic Bragg microcavity, demonstrating the design of tailored parametric resonances. Due to its generality and predictive power, our approach lays the foundation for the systematic study and engineering of structured PTCs, advancing the emerging field of space-time optics.
    
\end{abstract}
\maketitle
\section{INTRODUCTION} \label{sectionI}
Over the past several decades, progress in nanofabrication techniques and materials science have paved the way for engineered composites with unique electromagnetic properties. Examples include, but are not limited to metamaterials, metasurfaces, and photonic crystals constituted of dielectric or plasmonic nanoparticles \cite{basharin2015dielectric,smith2004metamaterials,yu2014flat,joannopoulos1997photonic,canos2021theory,kim2014optical}. These advanced materials are made possible by the ability to spatially structure matter at the nanoscale, inducing resonances that allow us to confine light and control its propagation.
This concept underpins their counterintuitive optical response and sets them apart from naturally occurring materials. 

In the quest to expand the available degrees of freedom for the control of light-matter interactions, the temporal dimension is currently being actively explored \cite{galiffi2022photonics,wang2024expanding,sounas2017non,yin2022floquet,ptitcyn2023tutorial,mostafa2024temporal,asgari2024photonic}. In principle, a spatially homogeneous material whose permittivity is modulated in time can behave as the temporal analog of a spatially engineered structure. Moreover, since the system is no longer passive, the idea brings up a host of previously unexplored opportunities for wave manipulation, such as magnetless nonreciprocity \cite{sounas2017non}, linear frequency generation \cite{galiffi2022photonics}, efficient surface wave excitation \cite{galiffi2020wood}, space-time Fresnel prisms \cite{li2023space}, and for overcoming bandwidth and absorption bounds of passive devices \cite{ptitcyn2023tutorial,hayran2024beyond}.

Among the exciting prospects, photonic time crystals (PTCs) \cite{ptitcyn2023tutorial,wang2024expanding,asgari2024photonic,saha2023photonic,lyubarov2022amplified} are arguably one of the most intriguing. They correspond to a special case of a time-modulated material, where the modulation is periodic, and thus obeys Floquet theorem \cite{asgari2024photonic}. In stark contrast with conventional photonic crystals, which are typically characterized by an energy bandgap where no wave propagation is allowed, momentum conservation together with the Floquet theorem give rise to a momentum bandgap~\cite{asgari2024photonic}. Inside the bandgap, Maxwell's equations provide a solution with exponentially increasing amplitude over time. Such an exotic response has been predicted to amplify spontaneous emission of an atom \cite{lyubarov2022amplified,park2022comment}, Cerenkov radiation from an electron \cite{dikopoltsev2022light} or an optical pulse with momentum residing in the gap \cite{lustig2023photonic}.

However, implementing PTCs in the lab provides a formidable challenge. Firstly, fast modulations in the order of twice the light frequency are needed to enter the bandgap. Secondly, large refractive index changes are required, with relative changes comparable to unity \cite{saha2023photonic}. Thirdly, realistic implementations of PTCs will require the modulation to be imparted on a finite structure, such as a material slab, a metasurface, or micro/nanoresonators. In general, such structures do not necessarily conserve momentum, and as a result constitute a different class of devices, with a phenomenology that starkly differs from that of PTCs in bulk media. To differentiate this general class of finite-size PTCs from time-modulated homogeneous media, we call the latter \emph{bulk} PTCs, and the former \emph{structured} PTCs. While most existing theoretical work has focused on bulk PTCs \cite{lyubarov2022amplified,koutserimpas2018,wang2020nonreciprocity}, studies on structured PTCs remain scarce.

The electromagnetic response of structured PTCs is primarily governed by resonances \cite{zurita2010resonances,ptitcyn2023floquet,stefanou2021light,panagiotidis2022inelastic,horsley2023eigenpulses,taravati2017nonreciprocal,vazquez2023incandescent}. Resonant behavior can strongly decrease the modulation thresholds for the observation of light amplification \cite{khurgin2024energy,wang2024expanding}. Hence, there is an urgent need for a rigorous theory of resonances, that can guide the design of realistic photonic time crystals in the optical regime.

In the absence of time modulation, it is well known that resonances in static structures are directly related to the resonant states (or quasinormal modes), which are source-free solutions of Maxwell's equations for outgoing waves \cite{lalanne2018light,both2021resonant,muljarov2018resonant,kristensen2020modeling,alpeggiani2017quasinormal}. Due to passivity constraints, they occur only at complex frequencies or in degenerate zero-pole pairs. Interestingly, far away from the resonator, their fields grow exponentially with distance, which has posed significant challenges for their normalization \cite{sauvan2022normalization,both2021resonant}. Once normalized, they serve as ``building blocks" for synthesizing the optical response at real frequencies \cite{weiss2018calculate,alpeggiani2017quasinormal}. As a result, they bring a much deeper physical insight into a given problem in comparison to brute-force numerical simulations \cite{lalanne2018light,both2021resonant}. Nowadays, the theory of resonant states is attracting a lot of interest in the nanophotonics community, and has been adapted to describe a wealth of physical phenomena, even at the quantum optics level \cite{muljarov2018resonant,lalanne2018light,ren2021quasinormal,franke2019quantization,koshelev2018asymmetric,canos2024bianisotropic,yuen2024exact,canos2023superscattering,valero2023exceptional,both2022nanophotonic}. 

Since structured PTCs inherit their resonant character from their static counterparts, the resonant states of the modulated structure must also be intimately connected to the new physics at play. However, beyond a few studies with case-specific models \cite{minkov2017exact,wang2024expanding,ptitcyn2023floquet}, the physics of such resonant states remains largely uncharted. The main reason stems from the lack of established methods to retrieve the resonant states of photonic time crystals of arbitrary shape, in contrast to the resonant states of static resonators, which can be obtained with most commercial numerical solvers \cite{lalanne2018light,ge2016quasinormal}.

In this work, we formulate a general, quantitative theory of resonant states for photonic time crystals. The new theory allows us to obtain the eigenfrequencies and eigenfields of a time-modulated structure with the sole knowledge of the resonant states of its static counterpart, which can be easily recovered with conventional approaches. Unlike the current models, our theory is self-consistent and does not require any fitting parameters, nor is it limited to resonant states with small decay rates. 

\begin{figure*}[htb!] 
    \centering
    \includegraphics[width=0.85\linewidth]{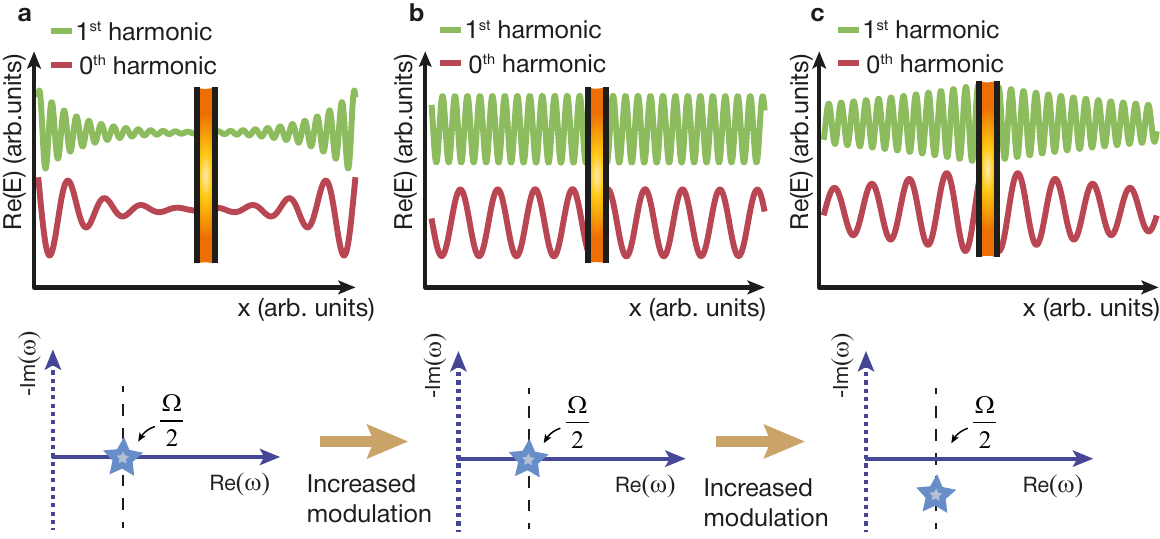}
    \caption{\label{fig1} Unusual behavior of resonant states in periodically time-modulated media. \textbf{a} Upper panel: 
    Fields of a resonant state of a dielectric slab with a resonance frequency near half the modulation frequency $\Omega$ and weak permittivity modulations. The slab is depicted schematically at the center of the plot. The fields display the known exponential divergence of resonant states in space, but unlike passive nanoresonators, they correspond to polychromatic sources that radiate several frequency harmonics. Lower panel: position of the eigenfrequency of the resonant state in the complex frequency plane, depicted with a star. \textbf{b-c} Same as \textbf{a}, but with increasingly large permittivity modulations. In \textbf{b}, the eigenfrequency becomes
    purely real and the resonant state does not diverge anymore, despite radiating to the far field. In \textbf{c}, the eigenfrequency has crossed the real axis and has an opposite sign of the imaginary part. As a result, the fields decay exponentially in space.}
\end{figure*}

Owing to its simplicity, our formulation allows to quantitatively describe several unusual properties of the time-modulated resonant states. To illustrate some of these, consider the example of a resonant state of a time-modulated dielectric slab, as shown in Fig.~\ref{fig1}. Unlike conventional static systems, the resonant states of structured PTCs radiate polychromatic fields in sidebands (harmonics) of the eigenfrequency $\tilde{\om}$, dictated by the modulation frequency $\Om$ [Fig.~\ref{fig1}(a)]. When the condition $\Om = 2\text{Re}(\tilde{\om})$ is fulfilled, a sufficiently strong modulation can bring the eigenfrequency to the real axis, resulting in a parametric resonance \cite{zurita2010resonances,globosits2024pseudounitary}, as depicted in the lower panel of Fig.~{\ref{fig1}}(b). Contrary to their static counterparts, the fields of ``parametric" resonant states still radiate, but do not grow exponentially away from the resonator [upper panel in Fig.~\ref{fig1}(b)]. Increasing the modulation amplitude further can induce a change in the sign of the imaginary part of $\tilde{\om}$, resulting in decaying eigenfields [Fig.~\ref{fig1}(c)]. This indicates the possibility of lasing upon exciting the resonant state, emulating a resonator with gain \cite{ren2021quasinormal}. 

Our theory provides insight into these and other intriguing phenomena. We discover that, for an arbitrary modulation, the original resonant states spawn an infinite number of ``replicas", which correspond to new resonant states of the modulated system. Moreover, when the original states are spectrally separated, to lowest order, all the eigenfrequencies of the modulated resonant states follow a universal quadratic dependence with the modulation amplitude. We also reveal the physical mechanism responsible for the formation of parametric resonances. These results are verified with a semi-analytical solution for a time-modulated dielectric slab, and subsequently applied for the design of a parametric resonance in a periodically-modulated Bragg microcavity.

\section{THEORETICAL SETTING} \label{sectionII}
\subsection{Static resonant states}
In the absence of sources, the curl Maxwell's equations in time domain take the form:
\begin{align}
    \nabla\times\mb{H}(t) = \frac{1}{c}\frac{\partial \mb{D}(t)}{\partial t}, \label{max1}\\
    \nabla\times\mb{E}(t) =-\frac{1}{c}\frac{\partial \mb{B}(t)}{\partial t}, \label{max2}
\end{align}
where $c$ is the speed of light in vacuum. To start, let us consider an unmodulated, linear optical system.  Taking into account the constitutive relations for a linear, causal medium \cite{muljarov2018resonant}, we write the curl Maxwell's equations in a compact notation as:
\begin{equation} \label{maxwell_time_compact}
    \ste{D}\mathbb{F}(t) = \frac{i}{c}\frac{\partial}{\partial t}\int \ste{P}(t-t')\mathbb{F}(t')\text{d}t'.
\end{equation}
In the above, the material kernel $\ste{P}(t-t')$ expresses the material response at time $t$ induced by a field at time $t'$. It is defined in terms of the permittivity and permeability tensors $\te{\eps},\te{\mu}$ and the bi-anisotropy tensors $\te{\xi},\te{\zeta}$: 
\begin{equation}
	\ste{P}(t) = 
	\begin{bmatrix}
		\te{\eps}(t)&\te{\xi}(t)\\\te{\zeta}(t)&\te{\mu}(t)
	\end{bmatrix}.
\end{equation} 
The operator $\ste{D}$ and the field vector $\mathbb{F}(t)$ are given by
\begin{equation}
    \ste{D} = \begin{pmatrix}
		\mb{0}& \nabla\times\\\nabla\times&\mb{0}
	\end{pmatrix},\hspace{0.25cm}
    \mathbb{F}(t) = \begin{bmatrix}
		\mb{E}(t)\\i\mb{H}(t)
	\end{bmatrix}.
\end{equation}
We emphasize that the fields $\mathbf{E}(t), \mathbf{H}(t)$ are the real, observable fields, and the imaginary unit on the right-hand-side of Eq.~(\ref{maxwell_time_compact}) is introduced for convenience. Furthermore, unless explicitly stated otherwise, we will consider all fields and material parameters to be spatially dependent. For brevity, we omit the spatial dependence in the notation, unless it is required for clarity.

Assuming a time dependence of the form $\exp(-i\om t)$, where $\om$ is the angular frequency, Eq.~(\ref{maxwell_time_compact}) reads \cite{muljarov2018resonant,lalanne2018light,both2021resonant,weiss2018calculate}: 
\begin{equation} \label{maxwell_basic}
    \frac{\om}{c}\ste{P}(\om)\mathbb{F}(\om)- \ste{D}\mathbb{F}(\om)= 0.
\end{equation}

We are interested in the solutions of Eq.~(\ref{maxwell_basic}) that behave as outgoing waves at large distances (they can radiate away). Typically, in passive systems, such solutions exist at complex frequencies \cite{alpeggiani2017quasinormal,lalanne2018light,both2021resonant,kristensen2020modeling}. These are the resonant states, eigensolutions of Eq.~(\ref{maxwell_basic}), with fields $\mathbb{F}_{\alpha}$ and eigenfrequencies $\tilde{\om}_{\alpha}$. Physically, the real part of the latter indicates their resonance frequency, and their imaginary part corresponds to the mode decay rate.

It is less well known that, as shown in Appendix~\ref{AppA}, for every resonant state $\alpha$ defined by the triplet $(\tilde{\om}_{\alpha},\mathbf{E}_{\alpha},\mathbf{H}_{\alpha})$ there exists another, which we call a \emph{negative twin} and label as ``$-\alpha$''. A negative twin $-\alpha$ is characterized by the fact that its eigenfrequency and eigenfields are related to those of the resonant state $\alpha$ as $(\tilde{\om}_{-\alpha},\mathbf{E}_{-\alpha},\mathbf{H}_{-\alpha}) = (-\tilde{\om}^*_{\alpha},\mathbf{E}_{\alpha}^*,\mathbf{H}_{\alpha}^*)$. Crucially, this means that their eigenfrequencies have \emph{negative} real parts, unlike conventional resonant states. While the influence of the negative twins can be usually neglected in passive systems~\cite{lalanne2018light,kristensen2020modeling}, they are of prime importance to describe the physics of time-modulated resonators, as will be discussed in the last sections.

\subsection{Maxwell's equations of a structured PTC}
If the material is modulated in time, Maxwell's equations can still be cast in the frequency domain, at the cost of an additional convolution in frequency \cite{ptitcyn2023floquet}. Accounting for the explicit time dependence of the material kernel, Maxwell's equations take the following form, generalizing Eq.~(\ref{maxwell_time_compact}) :
\begin{equation} \label{max3}
    \ste{D}\mathbb{F}(t) = \frac{i}{c}\frac{\partial}{\partial t}\int \ste{P}(t,t-t')\mathbb{F}(t')\text{d}t'.
\end{equation}
 The time variable $t$ now expresses the explicit time dependence of the material properties. After performing a double Fourier transform with respect to $t$ and $t'$, the frequency domain version of Eq.~(\ref{max3}) reads (refer to Appendix~\ref{AppB} for a detailed derivation): 
\begin{equation} \label{maxwell_compact}
	\frac{\om}{c}\int \ste{P}(\om-\om',\om')\mathbb{F}(\om')\text{d}\om'- \ste{D}\mathbb{F}(\om)= 0.
\end{equation}
Equation (\ref{maxwell_compact}), must be satisfied by all the resonant states in such systems, which in addition also fulfill outgoing radiation conditions. Similarly to passive resonators, their eigenfrequencies $\tilde{\omega}_{\alpha}$ will typically be complex.

Next, we particularize Eq.~(\ref{maxwell_compact}) to time-periodic media. Consider the system is modulated periodically with a frequency $\Om$. Then, the kernel $\ste{P}(t,t-t')$ in Eq.~(\ref{max3}) can be expanded in a discrete series of time harmonics:
\begin{equation} \label{Ptt}
    \ste{P}(t,t-t') = (2\pi)^{-1/2}\sum_n\ste{P}_n(t-t')e^{-i n\Om t},
\end{equation}
where $n$ runs over all integers, and the kernel $\ste{P}_n(t-t')$ denotes the material tensor associated with the time harmonic $n$. As shown in Appendix~\ref{AppC}, the kernel $\ste{P}(\om-\om',\om')$ takes the form $\ste{P}(\om-\om',\om') = \sum_n \ste{P}_n(\om')\delta(\om-\om'-n\Om)$ [Eq.~(\ref{FTptt3})], where $\delta(\om-\om'-n\Om)$ is a delta distribution. As a result, after the manipulations detailed in Appendix~\ref{AppC}, the integral convolution in Eq.~(\ref{maxwell_compact}) simplifies to a summation over all the discrete harmonics. We then arrive at the homogeneous Maxwell's equations of a generic structured PTC:

    \begin{equation} \label{maxwell_ptc}
        \frac{\om_m}{c}\sum_n\ste{P}_{m-n}(\om_n)\mathbb{F}_n- \ste{D}\mathbb{F}_m= 0.
    \end{equation}
In the above, we have defined $\om_m \equiv \om + m\Om$. $\mathbb{F}_m$ corresponds to the contribution of the field to the $m^{\text{th}}$ time harmonic. Physically, the latter can be understood as follows. Consider the system is driven by a continuous wave excitation at frequency $\om$. For a periodic material modulation with frequency $\Om$, the time-dependent electromagnetic fields will then be given by
\begin{equation}
    \mathbb{F}(t) = (2\pi)^{-1/2}\sum_m\text{Re}\left[\mathbb{F}_m (\om) e^{-i\om_m t}\right].
    \end{equation} 
Hence, each field component $\mathbb{F}_m(\om)$ corresponds to the field contribution oscillating at the shifted frequency $\om_m$. The $\om$ dependence of $\mathbb{F}_m(\om)$ will also be omitted from this point onward.


Furthermore, to avoid any confusion with the notation later on, time harmonics will always be denoted by latin subscripts $m,n$, and resonant states (both static and modulated), will be labelled with greek letters, when needed. 

Equation ~(\ref{maxwell_ptc}) is the starting point for the derivation of the resonant-state expansion, that we develop in the following.

\section{RESONANT-STATE EXPANSION OF STRUCTURED PTCS} \label{sectionIII}
\subsection{Formalism} \label{sectionIIIa}
We are interested in finding a rigorous procedure to obtain the eigenfrequencies and eigenfields of the resonant states in a time-periodic structure. In a first step, we rewrite \Eq{maxwell_ptc}, moving the terms caused by the modulation to the right-hand side:
\begin{align} \label{maxwell_inhom}
		\frac{\om_m}{c}\ste{P}_{0}(\om)\mathbb{F}_m- \ste{D}\mathbb{F}_{m} = \sum_{n\ne m}\mathbb{J}_{m,n}(\om),\\
		\mathbb{J}_{m,n}(\om) = -\frac{\om_m}{c}\ste{P}_{m-n}(\om_n)\mathbb{F}_n. \label{maxwell_inhom2}
\end{align}
An arbitrary periodic modulation of the material parameters might also induce a static shift of the material tensor. This can be straightforwardly accounted for in Eq.~(\ref{maxwell_inhom}) by including the $n=m$ term in the summation on the right-hand-side. However, perturbations to the static term are well studied~\cite{muljarov2011brillouin,muljarov2018resonant}. To reveal the physics introduced by pure temporal modulations, we will consider that no static shift takes place.

The left-hand side of Eq.~(\ref{maxwell_inhom}) corresponds to the homogeneous Maxwell’s equations of the static system [Eq.~(\ref{maxwell_basic})], with frequencies shifted by $m\Om$. Therefore, its solutions are simply the resonant states of the unmodulated structure, but oscillating at eigenfrequencies shifted by $m\Om$. We will show in Section~\ref{sectionIV} that these solutions give rise to new resonant states in periodically modulated resonators, which we term \emph{replicas}.

From Eq.~(\ref{maxwell_inhom}), we see that the periodic modulation can be regarded as a sum of current sources $\mathbb{J}_{m,n}$ ``perturbing" the solutions of the $m^{\text{th}}$ harmonic. As a result, the resonant states of different harmonics become coupled and can interact.

The solution of Eq.~(\ref{maxwell_inhom}) for an arbitrary current source can be obtained with the help of the Green's function, $\ste{G}(\om;\mb{r},\mb{r}')$ \cite{muljarov2018resonant}. Building on this result, the contribution of the eigenfield in the $m^{\text{th}}$ time harmonic, $\mathbb{F}_m$, can be written as: 
\begin{equation} \label{green_current}
	\mathbb{F}_m(\mb{r}) = \sum_{n\ne m}\int \ste{G}(\om_m;\mb{r},\mb{r}')\mathbb{J}_{m,n}(\om;\mb{r}')\text{d}\mb{r}'.
\end{equation}
In Eq.~(\ref{green_current}), the integral is only non-vanishing inside the resonator, since the time modulation, and hence the induced currents, are only non-zero there. In this scenario, the Green's function admits an expansion in terms of the resonant states of the static resonator \cite{muljarov2018resonant}:
\begin{equation} \label{green_unmod}
	\ste{G}(\om_m;\mb{r},\mb{r}') = c\sum_{\alpha}\frac{\mathbb{F}_{\alpha}(\mb{r})\otimes \mathbb{F}_{\alpha}(\mb{r'})}{\om_m-\tilde{\om}_{\alpha}}.
\end{equation}
In Eq.~(\ref{green_unmod}), the resonant states are assumed to be normalized with the norm given in Refs.~\cite{muljarov2018resonant,weiss2018calculate}. This expansion ensures the completeness of the resonant states inside the resonator, which is well established in the literature \cite{muljarov2011brillouin,lalanne2018light,weinstein1969open,kristensen2020modeling,both2021resonant}. Namely, any field inside can be expanded as a linear combination of resonant states \cite{both2021resonant,lalanne2018light}.

We can take advantage of this fact to expand the fields of the modulated resonant states in terms of the static ones. Plugging \Eq{green_unmod} in \Eq{green_current}, the field of a time-modulated resonant state at the time harmonic $m$ can then be written as: 
\begin{equation} \label{field_expansion}
    \mathbb{F}_m(\mb{r}) = \sum_{\alpha}c_{\alpha,m}\mathbb{F}_{\alpha}(\mb{r}).
\end{equation}
The coefficients $c_{\alpha,m}$ are now the expansion coefficients weighting the contribution of the static resonant state $\alpha$ to the modulated eigenfield in the time harmonic $m$. Their explicit expression can be found in Appendix~\ref{AppD}.

Following the steps detailed in Appendix~\ref{AppD}, we arrive at the eigenvalue problem for the eigenfrequencies of the time-modulated system, $\tilde{\om}$:
\begin{equation} \label{RSE}
	(\tilde{\om}+m\Om-\tilde{\om}_{\alpha})c_{\alpha,m} = -(\tilde{\om}+m\Om)\sum_{n\neq m,\beta}V_{\alpha,\beta}^{n,m}c_{\beta,n},
\end{equation}
with the matrix elements $V_{\alpha,\beta}^{n,m}$ given by:
\begin{equation} \label{Vab}
	V_{\alpha,\beta}^{n,m} = \int \mathbb{F}_{\alpha}(\mb{r})\cdot \ste{P}_{m-n}(\mb{r};\tilde{\om}+n\Om)\mathbb{F}_{\beta}(\mb{r})\text{d}\mb{r},
\end{equation}
where the integral runs over the volume affected by the modulation. In Eq.~(\ref{RSE}), we have made the replacement $\om\rightarrow\tilde{\om}$ to differentiate the new eigenfrequencies from any arbitrary frequency $\om$.

Equation~(\ref{RSE}) is the central result of this study. It allows reconstructing the eigenmodes of arbitrary, structured PTCs analytically, with the sole knowledge of the resonant states of the static system, and the Fourier transform of the material modulation.

We remark that if the material is dispersive, such as in the case of metals, the eigenvalue problem of Eq.~(\ref{RSE}) is nonlinear. However, it can be easily linearized with the method described in Ref.~\cite{muljarov2016resonant}.

We can now formalize some very general properties of the modulated resonant states from Eq.~(\ref{RSE}) and Eq.~(\ref{Vab}). In the first place, as could be expected, the matrix elements coupling different time harmonics will be generally nonzero. Therefore, the modulation can potentially form polychromatic modes, which radiate at multiple frequencies. For instance, if an eigenmode couples to the $ m = -1, 0, 1 $ time harmonics, then the corresponding eigenfield components $ \mathbb{F}_m(\mathbf{r}) $ for these harmonics will be nonzero.

Secondly, we see that if the material modulation is homogeneous within the system [$\ste{P}_n(\mb{r};\om)=\ste{P}_n(\om)$], the modulation cannot hybridize eigenmodes with different spatial symmetries in the static resonator. Consider for example a resonator with a mirror plane. An eigenmode $\alpha$ that is even with respect to that mirror plane will not couple with a mode $\beta$ that is odd, since the corresponding $V_{\alpha,\beta}^{n,m}$ will vanish. Lastly, the interaction between two static resonant states will be maximized when the modulation is introduced in the regions where eigenmodes have the strongest fields. These are important design rules that have no analog in bulk time modulated media, but must be carefully considered upon conceiving structured PTCs.  

In most practical scenarios, the formalism can be simplified even further. For the vast majority of cases of interest, only the permittivity is modulated \cite{galiffi2022photonics,khurgin2024energy}. Moreover, the phenomenology of periodic time-varying media is typically more pronounced in the absence of material dispersion \cite{asadchy2022parametric,zurita2010resonances,zurita2009reflection,ptitcyn2023floquet}. Finally, we consider the modulation to be isotropic within the resonator. With these three assumptions, and denoting by $\varepsilon_m$ the $m^{\text{th}}$ Fourier coefficient of the permittivity, Eq.~(\ref{Vab}) reduces to
\begin{equation} \label{Vab_simp}
	V_{\alpha,\beta}^{n,m} = \int \mathbf{E}_{\alpha}(\mb{r})\cdot \varepsilon_{m-n}\mathbf{E}_{\beta}(\mb{r})\text{d}\mb{r}.
\end{equation}
As a result, the eigenvalue problem becomes linear.

For the sake of clarity, the rest of the article will mainly focus the discussion on the physics of structured PTCs fulfilling the assumptions above.

\subsection{Verification for a time-modulated dielectric slab}
\begin{figure*}[t] 
    \centering
    \includegraphics[width=0.85\linewidth]{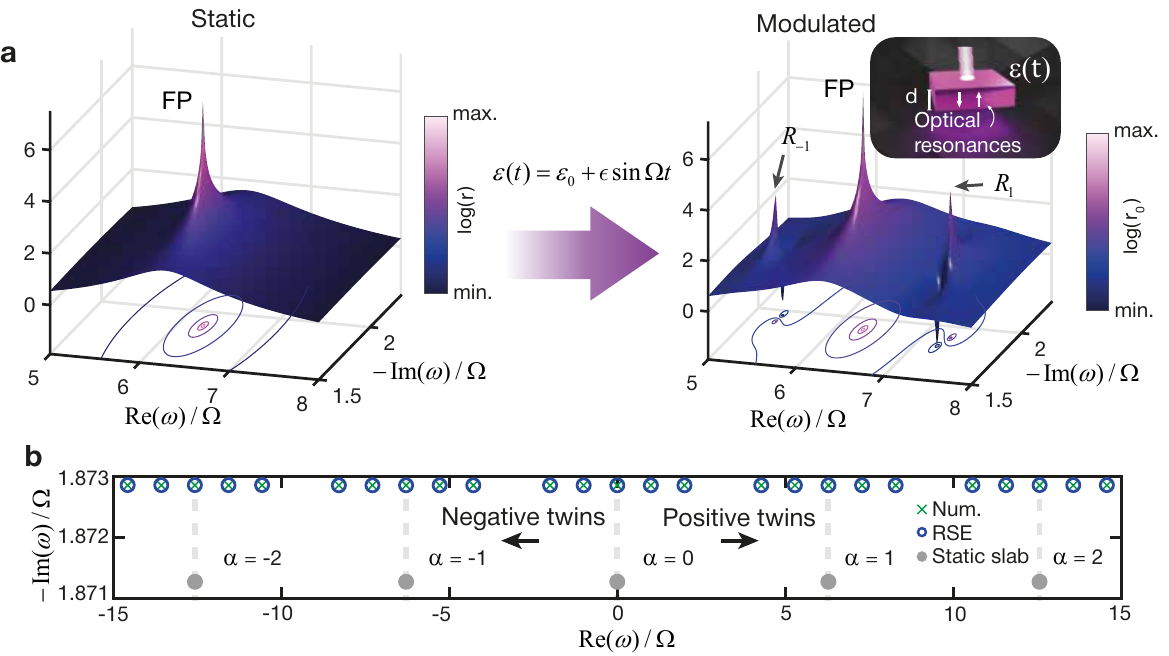}
    \caption{\label{fig2} Resonant states of a time-modulated dielectric slab. \textbf{a} Left panel: Reflection coefficient ($r$) in the complex plane for a static slab, in the vicinity of the Fabry-Perot resonance (FP) with $\alpha = 1$. The frequency is normalized by $\Omega$. The reflection diverges at the position of the poles. Right panel: reflection coefficient for the $0^{\text{th}}$ time harmonic ($r_0$), when the slab permittivity is sinusoidally modulated with a modulation amplitude $\epsilon = 0.8$. New poles, labeled $R_{\pm 1}$, emerge in the vicinity of the original Fabry-Perot resonance. Parameters used: Slab with normalized thickness $L_{\text{n}} = d\Om n_{\text{u}}/c =  0.5$, unmodulated permittivity $\varepsilon_{\text{u}} =n_{\text{u}}^2= 5.25$, and modulation frequency $\Omega = 300$ THz. $11$ time harmonics were used for the calculation. \textbf{b} Comparison between the numerically obtained eigenfrequencies and those obtained upon solving Eq.~(\ref{RSE}). Good visual agreement can be seen between the two approaches. The filled gray dots show the analytical solutions of the eigenfrequencies in the static slab. To solve Eq.~(\ref{RSE}), we retained  $2N_{\text{S}}+1=$15 eigenmodes in the expansion and $11$ time harmonics. Only the modes with the largest contributions to the $0^{\text{th}}$ time harmonic are shown.}
\end{figure*}

When a nanostructure is excited by an incoming wave, the resonant states manifest themselves as poles in the optical response, typically at complex frequencies. To gain preliminary insight into how these resonances evolve under temporal modulation, and validate our theoretical framework, it is instructive to examine the optical response of a simple structured PTC.  For this purpose, we consider a dielectric slab under normal-incidence plane wave excitation [sketched in the right panel of Fig.~\ref{fig2}(a)], whose static resonant states can be found analytically \cite{weiss2018calculate,muljarov2011brillouin}. 
Specifically, they correspond to evenly spaced Fabry-Perot resonances, with eigenfrequencies 
\begin{equation}
    \tilde{\om}_{\alpha} = \frac{\alpha \pi c}{dn_{\text{u}}} - i\gamma,
\end{equation}
where $d$ is the slab thickness, $n_{\text{u}}$ is the static refractive index of the slab (the subscript ``u'' stems from ``unmodulated''), $\gamma$ is the radiative decay rate and $\alpha$ is an integer number characterizing the resonant state. In the limit of small decay rates, $|\alpha|$ can be interpreted as the number of maxima in the standing wave forming the mode. 

The poles (eigenfrequencies), associated to the Fabry-Perot resonant states can be observed, for instance, in the diverging behavior of the reflected field amplitude $E_{\text{r}}$, as shown in the left panel of Fig.~\ref{fig2}(a) for the $\alpha = 1$ resonant state.

Next, we introduce a sinusoidal permittivity modulation of the form $\varepsilon(t) = \varepsilon_{\text{u}} + \epsilon \sin \Om t$, and choose, for illustration purposes, $\Om$ to be much smaller than the spectral distance between two neighboring Fabry-Perot resonant states (other situations will be studied later on). We consider strong modulation amplitudes, in the order of unity. 


The spectral response of the time-modulated dielectric slab can be efficiently obtained with the harmonic balance method \cite{zurita2009reflection,zurita2010resonances}. With this approach, we calculated the amplitude of the reflected field in the $0^{\text{th}}$ harmonic, $E_{\text{r,n}=0}$ [right panel in Fig.~\ref{fig2}(a)]. Notably, despite the strong modulation amplitude, the $\alpha = 1$ resonant state is barely affected. Instead, we observe the emergence of new peaks on each side of the original resonance, labelled $R_{\pm 1}$, suggesting the appearance of new resonant states. This dramatic difference between the static and modulated structures already hints at the exotic physics of the resonant states in such systems.

Using Eq.~(\ref{RSE}), together with Eq.~(\ref{Vab_simp}), we calculated the eigenfields and eigenfrequencies of the time-modulated structure. In principle, infinitely many harmonics $m$ and static resonant states $\alpha$ should be used for completeness. However, a sufficiently large truncated basis set with $m\in [-N,N]$, $\alpha \in [-N_{\text{S}},N_{\text{S}}]$ can already yield results with high accuracy, as demonstrated in the following.

The eigenfrequencies are displayed in Fig.~\ref{fig2}(b), and compared with an independent calculation by a pole-searching algorithm (refer to Appendix \ref{AppE} for details). Excellent visual agreement can be observed between the two approaches, highlighting the validity of the proposed formalism. For additional convergence tests, we refer the reader to section S1 of the Supplemental Material. 

Several interesting facts can already be drawn from Fig.~\ref{fig2}(b). Firstly, the reader can readily appreciate the negative twins of resonant states, distributed symmetrically with respect to their positive counterparts. The symmetry in the eigenspectrum is present in both modulated and static cases, as discussed in Appendix \ref{AppA}. Secondly, as could be expected from the behavior of the reflected field amplitude in Fig.~\ref{fig2}(a), the original Fabry-Perot modes are barely affected by the modulation. Thirdly, for each static mode, we observe a number of additional resonant states emerging, with eigenfrequencies approximately shifted by integer multiples of the modulation frequency, $\tilde{\omega}\approx\tilde{\omega}_{\alpha}+l\Om$, with $l\in \mathbb{Z}$. The eigenfrequencies $\tilde{\om}_{1}\pm \Om$ indeed agree well with the maxima of the side peaks observed in the right panel of Fig.~\ref{fig2}(a). We call these new modes \emph{replicas}. Their physical origin will be clarified in the next section. 
\section{RESONANT STATE PHYSICS FOR WEAK MODULATIONS} \label{sectionIV}
\begin{figure}[t] 
    \centering
    \includegraphics[width=0.75\linewidth]{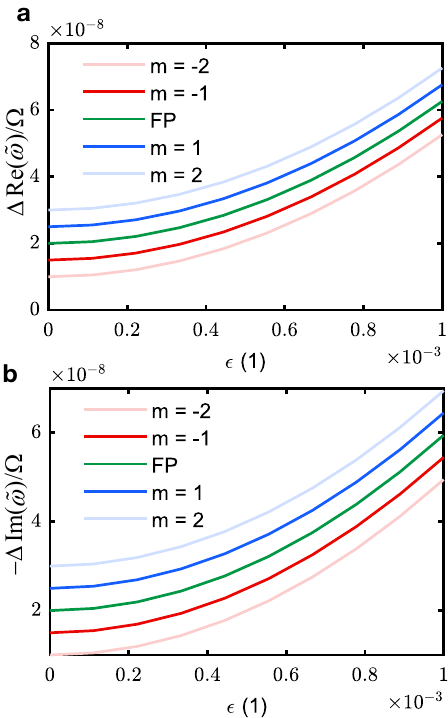}
    \caption{\label{fig3} Behavior of the resonant states for weak modulation amplitudes $\epsilon$. \textbf{a} Changes in the real part of the eigenfrequencies versus weak modulation amplitudes for the first ($\alpha = 1$), Fabry-Perot resonance (FP) and its replicas, labelled by $m$. Specifically, $\Delta\text{Re}(\tilde{\om})$ represents the difference between the real parts of the eigenfrequencies when $\epsilon \neq 0$ and $\epsilon = 0$.  \textbf{b} Same as \textbf{a}, for the imaginary parts. It can be seen that all eigenfrequencies scale quadratically with the modulation amplitude, as predicted by the perturbative analysis. The plots in (a)-(b) have been offset along the vertical axis for better visualization, with an offset given by $2 \cdot 10^{-9}$. System parameters: normalized slab thickness $L_{\text{n}}= d\Om n_{\text{u}}/c =$ 0.5, $\Omega = 300$ THz, $\varepsilon_{\text{u}} = n_{\text{u}}^2 = 5.25$. $11$ time harmonics were used for the calculation.}
\end{figure}


Having validated the quantitative correctness of our theory, we are now ready to use it for predicting the effect of a periodic time-modulation on the static resonant states of an arbitrary resonator. At first, we will focus on the influence of weak modulation amplitudes. In addition, we will consider only modulation frequencies smaller than the minimal separation between the two closest modes of the unmodulated structure, namely $\min\{|\tilde{\om}_{\alpha}-\tilde{\om}_{\beta}|\}\gg\Om$, for all $\beta\neq \alpha $. While these are apparently restrictive conditions, we will see that they can already reveal a number of interesting features. In particular, they enable us to solve Eq.~(\ref{RSE}) by an expansion of the matrix elements $V_{\alpha,\beta}^{n,m}$ in powers of $\epsilon$. In the spirit of perturbation theory \cite{sakurai2020modern}, we write a solution for the eigenfrequency $\tilde{\om}$ and the expansion coefficients $c_{\beta,n}$ of Eq.~(\ref{field_expansion}) for a given time-modulated mode in the form:
\begin{align} \label{pert_exp}
	\tilde{\om} \approx \tilde{\om}^{(0)} + \tilde{\om}^{(1)} + \tilde{\om}^{(2)}+\cdots,\\
	c_{\beta, n } \approx c_{\beta, n }^{(0)} + c_{\beta, n }^{(1)} + c_{\beta, n }^{(2)}+\cdots,\label{pert_exp2}
\end{align}
where the superscript indicates the related power of $\epsilon$ in the expansion.

The replica eigenmodes can be directly understood from the $0^{\text{th}}$ order approximation. In a way, this approximation is analogous to the empty ``temporal" lattice in condensed matter Floquet systems \cite{vogl2020effective,bukov2016schrieffer,d2014long}. Introducing Eqs.~(\ref{pert_exp}-\ref{pert_exp2}) in Eq.~(\ref{RSE}) and keeping only the $0^{\text{th}}$ order terms, we obtain:
\begin{align} 
	\tilde{\om}^{(0)} &= \tilde{\om}_{\alpha}-m\Om,\label{PT0}\\
	\mathbb{F}_m^{(0)} &= \mathbb{F}_{\alpha},\label{PT01}\\
    \mathbb{F}_n^{(0)} &= 0, \hspace{0.5cm} n\neq m.\label{PT02}
\end{align}

Remarkably, Eq.~(\ref{PT0}) tells us that upon weakly modulating an arbitrary resonator, every static resonant state $\alpha$ will spawn infinitely many replicas in the complex plane, with their eigenfrequencies shifted by integer multiples of $\Om$. 

Fig.~\ref{fig3}(a) shows the real parts of the eigenfrequencies of the time modulated dielectric slab considered in Fig.~\ref{fig2}, when $\epsilon \rightarrow 0$, but $\Om$ is finite. We indeed observe replicas for each of the Fabry-Perot modes (only shown up to $|m| = 2$ for better visualization).

In addition, Eqs.~(\ref{PT01}) and~(\ref{PT02}) show that the replica eigenfields are identical to the static resonant states, but their eigenfields only contribute to the $m^{\text{th}}$ time harmonic. Therefore, while replicas are true physical solutions of Maxwell's equations in the time-modulated structure, they seem at first glance not to contribute to the radiated fields upon plane wave illumination. 

This is because the $m^{\text{th}}$ replica can only couple to waves with frequencies $\om_m$, which are absent in the incident field. Therefore, while the zeroth order approximation provides an excellent guess for the spectral position of the resonances observed in Fig.~\ref{fig2}(a), higher order approximations are required to explain their contribution to the reflected fields in the $0^{\text{th}}$ harmonic.

Ultimately, replicas are an inherent result of coupling solutions of Maxwell's equations for different harmonics of the type given by Eq.~(\ref{green_current}), and they appear in a manner analogous to the Floquet band structure in bulk PTCs~\cite{asgari2024photonic,lyubarov2022amplified,zurita2009reflection}. However, they are a unique feature of structured PTCs, since they cannot be understood without knowledge of the static eigenmodes of the (spatially finite) structure.

We continue by investigating higher order terms in the perturbation series. Interestingly, unlike conventional static perturbations \cite{lalanne2018light,both2019first,both2021resonant,yang2015simple}, the first order term always vanishes, $\tilde{\om}^{(1)} = 0$. This is due to the summation on the right-hand side of Eq.~(\ref{RSE}) only running through harmonics $n\neq m$, as a result of the choice of the modulation in Eq.~(\ref{maxwell_inhom}) (see discussion in Section~\ref{sectionIIIa}).

Conversely, first-order corrections to the eigenvectors $c_{\beta, n}^{(1)}$ do not vanish. The eigenfields in the $n^{\text{th}}$ time harmonic can then be expressed as
\begin{equation} \label{fieldO1}
    \mathbb{F}_n \approx \mathbb{F}_{n}^{(0)} + \sum_{\beta,n}c_{\beta, n}^{(1)}\mathbb{F}_{\beta}.
\end{equation}

Equation~(\ref{fieldO1}) implies that any small modulation amplitude $\epsilon$ will already introduce mixing between the different harmonics. As a result, the original eigenmodes and their replicas become polychromatic. Replicas can then be excited by the incident field, which explains their contribution to the reflected field in the numerical results of Fig.~\ref{fig2}(a).

From the above, another important conclusion can be extracted: under weak, non-static periodic modulations of amplitude $\epsilon$, the eigenfrequencies of a resonant state scale quadratically as $\tilde{\om} = \tilde{\om}_{\alpha}-m\Om + C\epsilon^2$. $C$ is a constant that depends on the overlap integral between the fields of the static mode $\alpha$ and all the other resonant states. An explicit expression for $\tilde{\om}^{(2)}$ is derived in Section S2 of the Supplemental Material. 
Figure~\ref{fig3}(a) and Fig.~\ref{fig3}(b) show the change of the real and imaginary parts of the eigenfrequencies for the $\alpha = 1$ Fabry-Perot mode and its replicas versus small values of $\epsilon$, again calculated with Eq.~(\ref{RSE}). In all cases, the quadratic response can be readily observed. It is worth mentioning that the eigenfrequency changes are very weak both in their real and imaginary parts; in the order of $8\cdot 10^{-5}$ per unit of $\epsilon$. This explains why the static Fabry-Perot mode and its replicas in Fig.~\ref{fig2}(b) do not appear shifted upon a strong modulation.

In this section, we have derived perturbation formulas for the resonant states of structured PTCs. We have found that arbitrarily weak modulation amplitudes spawn an infinite set of replicas from the static eigenmodes. Introducing a strong enough modulation allows replicas to couple to incident waves and contribute significantly to the spectral response of the resonator. Moreover, we have discovered that, regardless of the temporal shape of the material modulation, the eigenfrequencies of all the modes scale quadratically with $\epsilon$.

However, as mentioned initially, the results are valid only if the unperturbed resonant states are spectrally separated. In section~\ref{sectionV}, we explore the failure of perturbation theory for nearly degenerate resonant states and uncover the role of the latter in enabling parametric amplification.

\section{ORIGIN OF PARAMETRIC RESONANCES IN STRUCTURED PTCs} \label{sectionV}
\begin{figure*}[t] 
    \centering
    \includegraphics[width=0.85\linewidth]{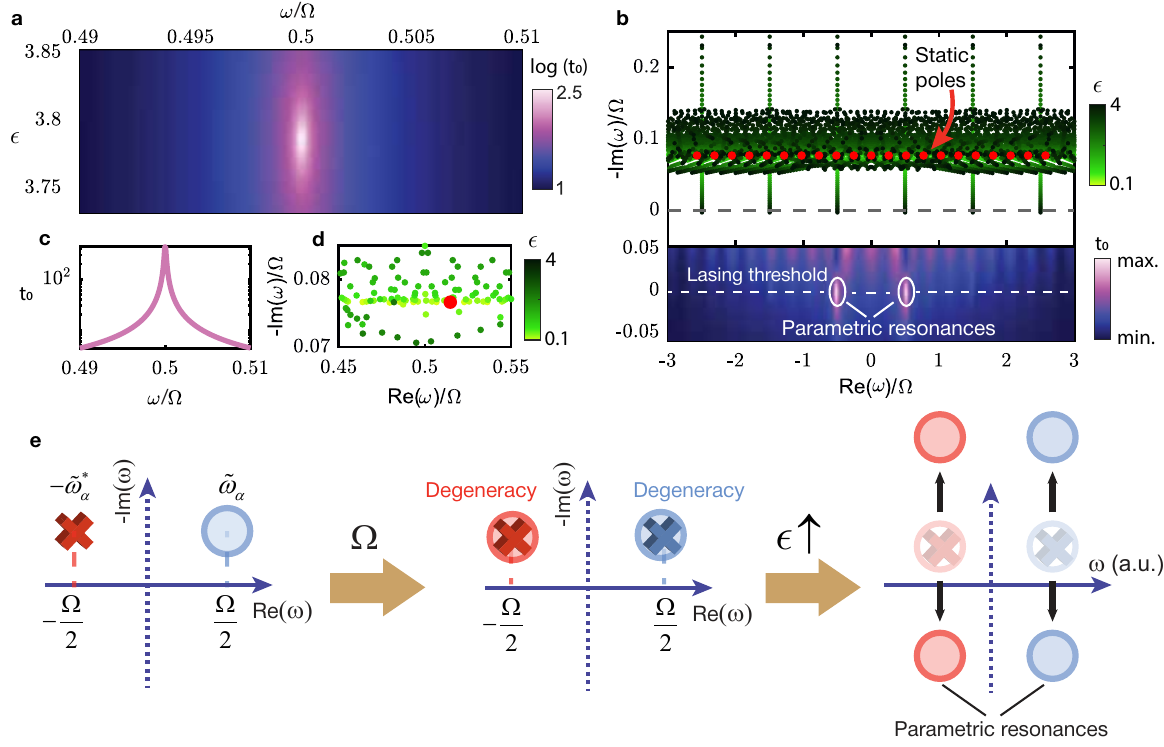}
    \caption{\label{fig4} Parametric resonances in a time-modulated dielectric slab. \textbf{a} Transmitted field amplitude in the $0^{\text{th}}$ time harmonic as a function of frequency and modulation amplitude, in the vicinity of a parametric resonance. $\text{t}_0$ is the transmission coefficient in the $0^{\text{th}}$ time harmonic. Modulation frequency $\Omega = 300$ THz. 21 time harmonics were used in all calculations. Slab with an unmodulated permittivity $\varepsilon_{\text{u}} = 5.25$ and normalized thickness $L_{\text{n}} = d\Om \sqrt{\varepsilon_{\text{u}}}/c = 12.2072$. \textbf{b} Upper panel: Behavior of the eigenfrequencies in the complex plane with increasing $\epsilon$, for the slab parameters in \textbf{a}. The resonant-state expansion is used with 81 resonant states and $21$ time harmonics. The eigenfrequencies of the static slab (red circles), are shown as a reference. Lower panel: Transmitted field amplitude of the $0^{\text{th}}$ harmonic in the complex plane for $\epsilon = 3.79$. The first two parametric resonances can be clearly seen in close proximity to the real axis. \textbf{c} Transmission coefficient in the $0^{\text{th}}$ time harmonic as a function of frequency, with $\epsilon = 3.79$. \textbf{d} Zoom of the eigenfrequencies in \textbf{b} near $\text{Re}(\om)/\Om = 0.5$. \textbf{e} Sketch of the mechanism responsible for the formation of parametric resonances. Leftmost panel displays a static resonant state and its negative twin. Central panel shows the replicas from the two modes upon introducing a modulation $\Om$, such that $\Om = 2\text{Re}(\tilde{\om}_{\alpha})$. Rightmost panel displays the modulated eigenfrequencies of the resonant states upon increasing the modulation amplitude.}
\end{figure*}

In a bulk PTC, an incident wave with frequency $\omega = \Omega/2$ undergoes parametric amplification (manifested as an exponential growth of the field amplitude in time), provided that the modulation amplitude is sufficiently large~\cite{galiffi2022photonics}. The analog situation in the case of a periodically modulated nanostructure is much less clear. In contrast, such systems can display \emph{parametric resonances}, corresponding to sharp peaks in the reflected and transmitted fields, with diverging amplitudes. Similar to bulk PTCs, they usually take place when the incident field is driven at $\om\approx \Om/2$. However, they only occur for some particular geometrical sizes and modulation amplitudes; they strongly depend on the structure design~\cite{zurita2010resonances,wang2024expanding,martinez2018parametric,asadchy2022parametric}. This suggests an intimate connection between parametric resonances and the resonant states of the time-modulated structure. In this section, we will shed light on this connection.

As a test case, we investigated a time-modulated dielectric slab, and optimize its thickness and modulation amplitude for the observation of a parametric resonance in the 0$^{\text{th}}$ time harmonic \cite{zurita2010resonances}. The results are shown in Fig.~\ref{fig4}(a).  It can be seen that the transmitted fields become strongly amplified in a narrow spectral range around $\om \approx \Om/2$, in the order of 1000 times the incident field amplitude. The transmitted field shows a resonant peak with a Lorentzian lineshape, as can be seen in Fig.~\ref{fig4}(c). We also remark that the resonance occurs only for very carefully selected thicknesses, confirming the fact that the amplification can no longer be described with the simple picture of a momentum bandgap, as in bulk PTCs \cite{galiffi2022photonics}.

We calculated the time-modulated resonant states with Eq.~(\ref{RSE}), and display the new eigenfrequencies in the top panel of Fig.~\ref{fig4}(b), as a function of $\epsilon$. The static eigenfrequencies of the slab are also shown as a reference. With any nonzero $\epsilon$, a large number of eigenfrequencies stemming from replicas are seen to appear in the vicinity of the static resonant states. A zoom-in for some of the eigenfrequencies in the cluttered region is displayed in Fig.~\ref{fig4}(d). These do not seem to play a significant role in the amplification. However, in Fig.~\ref{fig4}(b) we also observe pairs of eigenfrequencies that emerge from some of the unperturbed (static) eigenfrequencies and travel vertically along the imaginary axis in opposite directions. It can be seen that this happens near all static resonant states whose eigenfrequencies $\tilde{\om}$ have a real part at $\text{Re}(\tilde{\om}) \approx (m+1/2)\Om$, $m=(-N,\dots,N)$. What is more surprising, when $\epsilon$ is tuned to its value at the parametric resonance, several eigenfrequencies have zero imaginary part. Evidently, this has a drastic effect on the transmitted field, which can now diverge for real frequencies, explaining the amplification. The lower panel of Fig.~\ref{fig4}(b) indeed confirms that the position of the parametric resonances in the $0^{\text{th}}$ harmonic coincides with the eigenfrequencies of the first two lossless resonant states. Parametric resonances at larger frequencies (located at $3\Om/2,5\Om/2, \cdots$) can also display amplification in the $0^{\text{th}}$ time harmonic~\cite{zurita2010resonances}, but require very fine tuning of $\epsilon$ to be observed in the spectrum, up to 3 decimal digits.

Phenomenologically, the origin of parametric resonances is now clear: they are nothing more than resonant states whose (generally complex) eigenfrequencies are brought to the real axis upon a periodic modulation of the material. At first glance, the concept strongly resembles that of bound states in the continuum in passive systems~\cite{koshelev2018asymmetric,valero2023exceptional}. However, unlike the latter, parametric resonances can radiate into the far field even when the eigenfrequency is purely real.

Our goal now is to find a simplified model that captures the mechanism giving rise to this peculiar effect. In addition, we want to answer the following questions: (i) why do only static resonant states whose eigenfrequencies have a real part close to $\Om/2+m\Om$, $m=(-N,\dots,N)$ seem to yield parametric resonances? (ii) How can we choose (and design) resonant states that will display a parametric response for realistic modulation amplitudes? In what follows, we present the model and address the first question, providing insights into the role of static resonant states in parametric resonances. The answer to the second question will be discussed in Section \ref{sectionVI}. 

To keep the results general, we consider an arbitrary (static) resonator supporting a resonant state, labelled $\alpha$, and its negative twin, $-\alpha$, with eigenfrequencies $\tilde{\om}_{\alpha}$, and $\tilde{\om}_{-\alpha} = -\tilde{\om}^*_{\alpha}$, respectively.
Their positions in the complex plane are sketched in the leftmost panel of Fig.~\ref{fig4}(e). Inspired by the results in Fig.~\ref{fig4}(b), we introduce a periodic modulation such that $\Om = 2\text{Re}(\tilde{\om}_{\alpha})$. As discussed in Section \ref{sectionIV}, each resonant state spans a number of replicas. However, this particular choice of modulation frequency gives rise to a unique scenario, since the $m=-1$ replica of the negative twin becomes degenerate with the positive twin, as depicted in the middle panel of Fig.~\ref{fig4}(e). This is because using Eq.~(\ref{PT0}) we get $\tilde{\om}_{-\alpha}+\Om = \Om/2-i\gamma = \tilde{\om}_{\alpha}$. The same occurs for the eigenfrequency of the negative twin and the $m=1$ replica of the positive pair. Owing to their strong spectral overlap, these two eigenmodes play a dominant role in the formation of parametric resonances, as we show in the following.

Due to the degeneracy of the two resonant states, we are unable to use the perturbation formulas derived in Section \ref{sectionIV}, which are valid only when the spectral separation between the modes is large. Instead, we follow a different route and solve directly Eqs.~(\ref{RSE}), by assuming a sinusoidal modulation and keeping only the degenerate states in the mode basis. This leads to a reduced eigenvalue problem for the modulated eigenmodes:
\begin{equation} \label{PRsystem}
    \begin{pmatrix}
        \frac{\Om}{2}-i\gamma&&0\\
        -\Om V&&\frac{\Om}{2}-i\gamma
    \end{pmatrix}\mb{c} = \tilde{\om}\begin{pmatrix}
        1&&V\\
        -V&&1
    \end{pmatrix}\mb{c}.
\end{equation}
The top entry of the eigenvector $\mb{c}$ is the contribution of the static resonant state $\tilde{\om}_{\alpha}$, and the bottom entry is the contribution of the $m=-1$ replica of its negative twin. For a sinusoidal modulation, $V$ is derived from the coupling matrix elements in Eq.~(\ref{Vab_simp}) as:
\begin{equation}
    V = \frac{\epsilon}{2i}\int|\mb{E}_{\alpha}(\mb{r}')|^2d\mb{r}'\equiv\frac{\epsilon}{2i}I_{\alpha}.
\end{equation}
Unlike the general case in Eq.~(\ref{Vab_simp}), we see that the overlap integral between the mode fields is purely real. This is because, as explained in section \ref{sectionII} and Appendix \ref{AppA}, the field of the negative twin is just the complex conjugate of that of the positive one. 

Equation~(\ref{PRsystem}) can be solved exactly. However, it is more insightful to inspect the solution for small coupling coefficient, $V\ll 1$. With this last assumption, the eigenfrequencies can be found as:
\begin{equation} \label{dispPR}
        \tilde{\om}_{\pm} \approx \frac{\Om}{2}-i\gamma\mp iI_{\alpha}\frac{\epsilon}{4}\sqrt{\Om^2+4\gamma^2}. 
\end{equation}
The discriminant in Eq.~(\ref{dispPR}) is purely imaginary. Consequently, the new eigenfrequencies have the same real part $\text{Re}(\tilde{\om}_{\pm}) = \Om/2$ as the uncoupled eigenmodes, but for one of them the imaginary part increases as a function of $\epsilon$, while for the other one it decreases. An identical situation occurs for the coupled system of $-\alpha$ and the $m=1$ replica of $\alpha$. Both coupled sets of eigenfrequencies are sketched in the right-most panel of Fig.~\ref{fig4}(e).

Crucially, a parametric resonance takes place when the modulation amplitude is such that the time modulation provides enough energy to completely compensate the losses, bringing one of the hybrid eigenmodes to the real axis. From that point, increasing the modulation amplitude further can even change the sign of the imaginary part, introducing gain into the resonant state [see Fig.~\ref{fig1}(c)].

It is worth emphasizing that the strongest amplification will be observed when the eigenfrequency is near or at the real axis; introducing more gain or loss by increasing or decreasing $\epsilon$ will just dampen the resonance, as can be confirmed in the numerical results of Fig.~\ref{fig4}(a). 

We also clarify that the reduced model of Eq.~(\ref{PRsystem}) is an approximation, as it neglects the influence of all other static resonant states and their replicas. Their contributions can become significant for large $\epsilon$, which is required for compensating the losses of the Fabry-Perot modes of the dielectric slab. This explains the different growth rates with $\epsilon$ for the loss-like and gain-like resonant states in Fig.~\ref{fig4}(b). In addition, it can be expected that, in practice, the non-linearities in the material would force the parametric resonances to stay pinned on the real axis instead of crossing it, in a manner analogous to their effect on the behavior of lasers~\cite{esterhazy2014scalable,tureci2008ab}. 

To summarize, in this section we have investigated the link between resonant states and parametric resonances in a structured PTC. Parametric resonances are a consequence of resonant states whose eigenfrequencies are brought to the real axis by the resonant coupling induced by the modulation. This results in a strong amplification of the fields. Ultimately, its formation is mainly determined by the hybridization between a resonant state of the static system with the replica of its negative twin. Their coupling is maximized when the modulation frequency is around two times the real part of the static eigenfrequency. 

\section{DESIGNING A PARAMETRIC RESONANCE WITH A BRAGG MICROCAVITY} \label{sectionVI}
\begin{figure*}[t] 
    \centering
\includegraphics[width=0.85\linewidth]{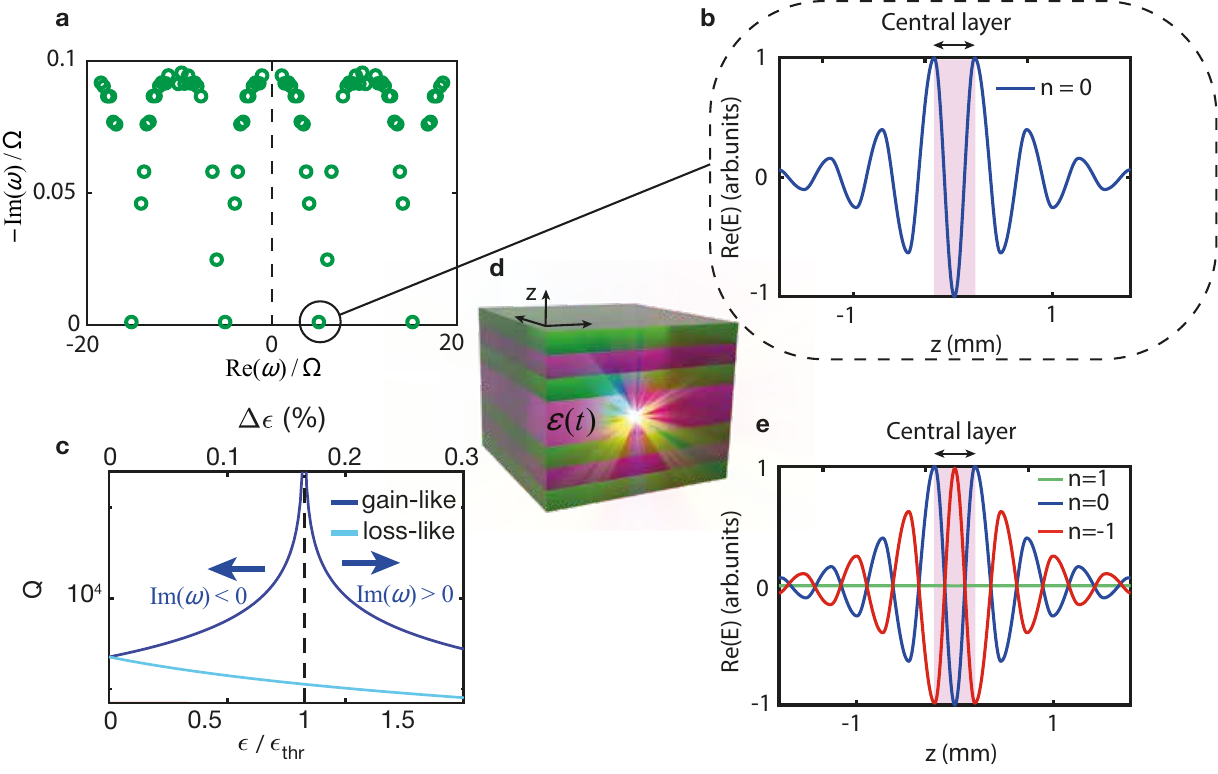}
\caption{\label{fig5} Design of a parametric resonance in a Bragg microcavity. \textbf{a} Spectrum of resonant states of the static microcavity used for the calculation. \textbf{b} Electric field distribution of the fundamental cavity mode. \textbf{c} Quality factor [$Q$, defined as $Q = \text{Re}(\tilde{\om})/2|\text{Im}(\tilde{\om})|$] of the eigenfrequency pairs emerging from the fundamental cavity mode vs. modulation amplitude $\epsilon$. The bottom horizontal axis shows $\epsilon$ normalized to the threshold value $\epsilon_{\text{thr}}$ obtained from Eq.~(\ref{PR_thr}). The horizontal axis at the top shows the modulation percentage, defined as $\Delta\epsilon = 100\epsilon/\varepsilon_1$, where $\varepsilon_1$ is the static permittivity of the central layer, given below. As indicated in the plot, the imaginary part of the gain-like resonant state is negative below the threshold, and positive above it. \textbf{d} Artistic representation of the Bragg microcavity under consideration and the parametric resonance, depicted as a polychromatic source with a hotspot at the center of the cavity (the number of layers is illustrative). The microcavity is embedded in vacuum, and consists of two Bragg mirrors sandwiching a cavity layer. Each Bragg mirror is made of 12 layers of alternating permittivities $\varepsilon_{1} = 10$ and $\varepsilon_2 = 4$, with thicknesses $l_{1} = 98\hspace{0.15cm}\text{nm}$ and $l_2 = 155\hspace{0.15cm}\text{nm}$. The cavity layer has a thickness of $l = 392$ nm and a time-varying permittivity $\varepsilon(t) = \varepsilon_1+\epsilon \sin(\Omega t)$. \textbf{e} Electric field of the parametric resonant state for the first harmonics ($n = -1,0,1$). The results in \textbf{c} and \textbf{e} were obtained with 21 time harmonics.}
\end{figure*}

Very recently, it has been predicted that light confinement in time-modulated nanoresonators could drastically reduce the modulation threshold necessary for light amplification, in comparison with bulk PTCs \cite{wang2024expanding,khurgin2024energy}. The analysis in Section \ref{sectionV} can now provide an intuitive understanding of why this is the case. From Eq.~(\ref{dispPR}), the modulation threshold $\epsilon_{\text{thr}}$ can be easily found by setting the imaginary part to zero, giving:
\begin{equation} \label{PR_thr}
    \epsilon_{\text{thr}} \approx \frac{4\gamma}{I_{\alpha}\sqrt{\Omega^2+4\gamma^2}} = \frac{2\gamma}{I_{\alpha}|\tilde{\om}_{\alpha}|}.
\end{equation}
Equation~(\ref{PR_thr}) is a novel result allowing, without any fitting parameters, to estimate the required modulation amplitude for any resonant state to become a parametric resonance. It can be seen that the larger the imaginary part of the eigenfrequency is (the larger the decay rate), the larger will $\epsilon_{\text{thr}}$ need to be to bring it to the real axis. Therefore, it becomes clear that eigenmodes with small $\gamma$ (large quality factors) can be more easily driven to the parametric regime. So far, however, practical designs implementing this idea are scarce and rely on complex metamaterial structures \cite{wang2024expanding}.  

Instead, to minimize structural complexity, we propose the realization of a parametric resonance with a standard planar semiconductor microcavity, sketched in Fig.~\ref{fig5}(d). The structure consists of two Bragg mirrors with alternating dielectric layers, surrounding a thicker dielectric layer, together forming the cavity. 

The eigenfrequencies of the unmodulated configuration are shown in Fig.~\ref{fig5}(a). A series of cavity modes with very small imaginary parts are seen to appear. Any of them is a suitable candidate to obtain a parametric resonance. For illustration, we choose the fundamental cavity mode, with an eigenfrequency denoted by $\tilde{\om}_{\text{f}}$. Its field distribution is shown in Fig.~\ref{fig5}(b). The latter has a maximum in the central layer (pink-shaded region), and its amplitude drops rapidly within the Bragg mirrors. As discussed in Section \ref{sectionIII}, the effect of the permittivity modulation is most pronounced in regions with the strongest modal field. Accordingly, we apply a sinusoidal modulation to the permittivity of the central layer. In addition, this configuration may allow for a simple experimental realization, because the material modulation can be imparted from the side of the microcavity \cite{rigneault1998resonant}, for instance by driving a nonlinearity of the material with a train of high intensity pulses.

Following the design rules from Section \ref{sectionV}, the system is then modulated at $\Om = 2\text{Re}(\tilde{\om}_{\text{f}})$. Interestingly, when $|\text{Im}(\tilde{\om}_{\text{f}})|\ll |\text{Re}(\tilde{\om}_{\text{f}})|$, and assuming small $\epsilon_{\text{thr}}$, Eq.~(\ref{PR_thr}) becomes $\epsilon_{\text{thr}} \approx (Q_{\text{f}}I_{\text{f}})^{-1}$, where $Q_{\text{f}} = |\text{Re}(\tilde{\om}_{\text{f}})/2\text{Im}(\tilde{\om}_{\text{f}})|$ is the quality factor of the unmodulated mode. For weak modulation amplitudes, this relation becomes very accurate, and can be used to quantitatively predict the threshold value $\epsilon_{\text{thr}}$ to observe a parametric resonance. It only requires the knowledge of the quality factor and the field of the static resonant state, which can be obtained analytically or with any standard numerical Maxwell solver \cite{multiphysics1998introduction}. 

Figure~\ref{fig5}(c) shows the evolution of the quality factor of the two hybrid resonant states spawning from the fundamental cavity mode with increasing $\epsilon$. The quality factor has been defined as $Q=|\text{Re}(\tilde{\om})/2\text{Im}(\tilde{\om})|$, since the imaginary part of the parametric resonant state changes sign upon crossing the real axis. It can be seen that the required permittivity modulation is only in the order of $0.16\%$ of the central layer permittivity, around 450 times less than the isolated slab. The bottom axis shows the ratio between the modulation amplitude $\epsilon$ and the threshold $\epsilon_{\text{thr}}$ calculated with Eq.~(\ref{PR_thr}). It can be confirmed that the parametric resonance indeed appears at the value predicted by our theory. As a side note, we remark that our definition of the quality factor, while similar to that of passive resonators \cite{lalanne2018light}, can no longer be associated to the radiation losses of the resonant state \cite{koshelev2018asymmetric,lalanne2018light}, since the parametric resonance is always strongly radiating. 

In contrast to its static counterpart, the eigenfields of the parametric resonant state are polychromatic [Fig.~\ref{fig5}(e)]. The $0^{\text{th}}$ harmonic retains the original spatial profile of the fundamental mode. Additionally, the resonant state now features a pronounced contribution in the $-1^{\text{st}}$ harmonic. Conversely, the contribution of the $1^{\text{st}}$ harmonic is negligible. This behavior confirms the hybridization of the static resonant state with the replica of its negative twin, as expected from the mechanism explained in Section \ref{sectionV}.

The results of this section underscore the potential of our general theory for the analysis and design of photonic space-time crystals. Specifically, by selecting a suitable high quality factor resonant state in a Bragg microcavity, we are able to achieve a parametric resonance with significantly reduced modulation threshold, compared to bulk PTCs. The device could be implemented with standard semiconductors, such as AlAs/GaAs \cite{stanley1994ultrahigh}. In addition, unlike the previous proposal based on a complex metamaterial \cite{wang2024expanding}, our structure presents the advantage of being easily grown via well-established molecular beam epitaxy techniques \cite{stanley1994ultrahigh,megahd2022planar}.

\section{DISCUSSION AND CONCLUSION} \label{sectionVII}

In this work, we have developed a quantitative theory to calculate the resonant states of structured PTCs of arbitrary shape. Our formalism only requires the knowledge of the resonant states of the static system, which can be obtained analytically or numerically with any commercial Maxwell solver, avoiding the need of resource-consuming simulations of the full problem. Moreover, unlike earlier attempts based on temporal coupled mode theory or sheet transition conditions \cite{asadchy2022parametric,karl2020frequency,minkov2017exact,rizza2024harnessing}, our theory does not introduce any limitation in terms of the structure size, number of modes, mode losses or spectral isolation, and does not require any fitting parameters. Our results were verified with a semi-analytical solution for a time-modulated dielectric slab.

The simplicity of our formulation has allowed us to obtain significant insights into the physics of these exotic eigenmodes. We have shown that any modulation will generate replicas of the original resonant states, which can then manifest as additional eigenfrequencies in the complex spectrum of observables. We have derived perturbation formulas to understand the behavior of the modulated resonant states for weak modulations, and revealed their polychromatic nature. Interestingly, if the resonant states are spectrally separated, all the eigenfrequencies scale quadratically with the modulation amplitude. This is a general phenomenology that occurs for all resonator shapes and sizes.

Unlike in passive resonators, the negative twins of the static resonant states have an essential role in the physics at play. Specifically, we have found that parametric amplification occurs mainly due to the hybridization of a resonant state and a replica of its negative twin. Their coupling generates an eigenmode that can reach, and even cross (in the absence of non-linearities), the real axis, forming a parametric resonance. This mechanism is qualitatively different from the parametric amplification mediated by a momentum bandgap in bulk PTCs, and should be carefully taken into account in the design of such devices.

Given the growing interest in time-modulated resonators and metasurfaces \cite{galiffi2022photonics,ptitcyn2023floquet,zubyuk2022externally,yang2024ultrafast,berte2024emergent,shilkin2024ultrafast,yin2022floquet,asgari2024photonic,rizza2024harnessing}, the theory presented here is expected to find broad application at the forefront of fundamental research in micro and nanophotonics. The number of platforms for exploration is extensive. For instance, our theory could be directly used to study the hybridization of polaritonic materials and structured PTCs. It could also shed light on the role of resonant phenomena in structures composed of epsilon-near-zero materials, which are being actively investigated for the realization of temporal reflection and refraction in the optical regime \cite{lustig2023time}. Furthermore, it can guide the design of devices for the realization of photon acceleration at the nanoscale \cite{shcherbakov2019photon}, parametric Huygens metasurfaces \cite{liu2018huygens,asadchy2022parametric}, or temporal Wood anomalies \cite{galiffi2020wood}. Crucially, it can be used to pave the way towards the experimental realization of parametric resonances in the optical regime \cite{wang2024expanding,dikopoltsev2022light,lyubarov2022amplified}.

\section*{ACKNOWLEDGMENTS}
The authors acknowledge helpful discussions with Jakob Hüpfl. This research was funded in whole, or in part, by the Austrian Science Fund (FWF) [Grants 10.55776/P36864 and 10.55776/PIN7240924]. For open access purposes, the authors have applied a CC BY public copyright license to any author-accepted manuscript version arising from this submission. ACV acknowledges funding by the project No 1.1.1.9/LZP/1/24/101 : "Non-Hermitian physics of spatiotemporal photonic crystals of arbitrary shape (PROTOTYPE)".

\appendix
\setlength{\parskip}{0.5em} 
\section{PROOF OF THE EXISTENCE OF NEGATIVE TWINS OF RESONANT STATES} \label{AppA}

Let us consider a particular resonant state $(\tilde{\omega}_{\alpha},\mathbb{F}_{\alpha})$ solution of Eq.~(\ref{maxwell_basic}). The negative twins arise naturally upon making the substitution $\tilde{\om}_{\alpha}\rightarrow-\tilde{\om}_{\alpha}^*$:
\begin{equation} \label{maxwell_conj}
    -\frac{\tilde{\om}_{\alpha}^*}{c}\ste{P}(\mb{r};-\tilde{\om}_{\alpha}^*)\mathbb{F}_{\alpha}(\mb{r};-\tilde{\om}_{\alpha}^*)- \ste{D}\mathbb{F}_{\alpha}(\mb{r};-\tilde{\om}_{\alpha}^*)= 0.
\end{equation}
The Fourier transform $\mathcal{F}(\om)$ of an arbitrary real signal $f(t)$ must satisfy Hermitian symmetry, $\mathcal{F}(-\om^*) = \mathcal{F}^*(\om)$ \cite{landau2013electrodynamics}. Imposing this constraint to the electric and magnetic fields yields:
\begin{equation} \label{maxwell_neg_pair}
    -\frac{\tilde{\om}_{\alpha}^*}{c}\ste{P}(\mb{r};-\tilde{\om}_{\alpha}^*)\mathbb{F}_{\alpha}^{\text{N}}(\mb{r};\tilde{\om}_{\alpha})- \ste{D}\mathbb{F}_{\alpha}^{\text{N}}(\mb{r};\tilde{\om}_{\alpha})= 0,
\end{equation}
with $\mathbb{F}_{\alpha}^{\text{N}}(\mb{r};\tilde{\om}_{\alpha})=\left[\mb{E}^*_{\alpha}(\mb{r};\tilde{\om}_{\alpha}),i\mb{H}^*_{\alpha}(\mb{r};\tilde{\om}_{\alpha})\right]^\text{T}$. Equation ~(\ref{maxwell_neg_pair}) shows that, to every resonant state fulfilling Eq.~(\ref{maxwell_basic}) at eigenfrequency $\tilde{\om}_{\alpha}$, we can associate a negative twin that also satisfies it at an eigenfrequency $-\tilde{\om}_{\alpha}^*$, with fields that are the complex-conjugates of the original resonant state. 

Furthermore, since the proof relies solely on general properties of the Fourier transform of the fields, it also applies to the resonant states of arbitrary time modulated resonators, defined by Eq.~(\ref{maxwell_compact}).

\section{DERIVATION OF EQUATION~(\ref{maxwell_compact})} \label{AppB}
Starting from Eq.~(\ref{max3}), in what follows we provide a detailed derivation of Eq.~(\ref{maxwell_compact}). We start by recalling the definitions of the Fourier transform of a function $f(t)$ and its inverse \cite{arfken2011mathematical}:
\begin{align} 
    f(\om) \equiv (2\pi)^{-1/2}\int f(t)e^{i\om t}\text{d}t,\label{FT}\\
    f(t) \equiv (2\pi)^{-1/2}\int f(\om)e^{-i\om t}\text{d}\om\label{invFT}.
\end{align}
Since the material tensor in Eq.~(\ref{max3}) depends on both $t$ and $t'$, its Fourier decomposition can be obtained by applying Eq.~(\ref{invFT}) twice to $\ste{P}(t,t')$, giving:
\begin{equation} \label{FTP}
    \ste{P}(t,t') = \frac{1}{2\pi}\iint \ste{P}(\om',\om)e^{-i\om t'}e^{-i\om't}\text{d}\om'\text{d}\om     
\end{equation}
Using Eq.~(\ref{FTP}) and the Fourier decomposition of $\mathbb{F}(t)$ we can rewrite Eq.~(\ref{max3}) as
\begin{equation} \label{max4}
    \begin{split}
    \int \ste{D}\mathbb{F}(\omega)e^{-i\omega t} \, \text{d}\omega 
    &= (2\pi)^{-1/2} \frac{i}{c} \frac{\partial}{\partial t} 
    \iiint  e^{-i\omega' t}\ste{P}(\omega', \omega) \\
    &\quad \times \mathbb{F}(t') e^{-i\omega (t-t')} \, \text{d}\omega' \, \text{d}\omega\text{d}t'.
    \end{split}
\end{equation}

Upon integrating over $t'$ we get, due to Eq.~(\ref{FT}):
\begin{equation} \label{max5}
    \begin{split}
    \int \ste{D}\mathbb{F}(\omega)e^{-i\omega t}\text{d}\omega  &= \frac{i}{c} \frac{\partial}{\partial t} 
    \iint  \ste{P}(\omega', \omega)e^{-i(\om+\om') t} \\
    &\quad \times \mathbb{F}(\om)\text{d}\omega'\text{d}\omega.
    \end{split}
\end{equation}
Making the change of variables $\om+\om'\rightarrow \om'$ inside the integral over $\om'$ in Eq.~(\ref{max5}) yields:
\begin{equation} \label{max6}
    \begin{split}
    \int{\ste{D}\mathbb{F}(\omega)e^{-i\omega t}\,\text{d}\omega} &= 
    \frac{i}{c}\frac{\partial}{\partial t}\int{e^{-i\omega't}} \\
    &\quad \times \int{\ste{P}(\omega'-\omega,\omega)\mathbb{F}(\omega)\,\text{d}\omega'\text{d}\omega}.
    \end{split}
    \end{equation}
    
Upon differentiating the right-hand side and making the variable exchange $\om\leftrightarrow\om'$, we readily obtain Eq.~(\ref{maxwell_compact}):
\begin{equation} \label{maxwell_compact_app}
    \frac{\om}{c}\int \ste{P}(\om-\om',\om')\mathbb{F}(\om')\text{d}\om'- \ste{D}\mathbb{F}(\om)= 0.
\end{equation}

\section{MAXWELL'S EQUATIONS FOR TIME-PERIODIC MEDIA} \label{AppC}
In this Appendix, we provide a detailed proof of Eq.~(\ref{maxwell_ptc}), which is a particularized version of Eq.~(\ref{maxwell_compact}) for the special case of time-periodic media. For clarity, we repeat here the general form of a material tensor modulated periodically in time, Eq.~(\ref{Ptt}):
\begin{equation} \label{ptt}
    \ste{P}(t,t-t') = (2\pi)^{-1/2}\sum_n\ste{P}_n(t-t')e^{-i n\Om t}.
\end{equation}
Using Eq.~(\ref{FT}), $\ste{P}(\om,\om')$ is given by
\begin{equation} \label{FTptt}
    \ste{P}(\om',\om) = \frac{1}{2\pi}\iint \ste{P}(t,t')e^{i\om t'}e^{i\om't}\text{d}t\text{d}t' .
\end{equation}
However, we are interested in a relation between the kernels $\ste{P}(t,t')$ and $\ste{P}(\om-\om',\om')$. For that purpose, we perform the change of variables $\om'\rightarrow \om-\om'$ and $\om\rightarrow \om'$ in Eq.~(\ref{FTptt}), obtaining:
\begin{equation} \label{FTptt2}
    \begin{split}
        \ste{P}(\omega - \omega', \omega') &= \frac{1}{2\pi} \iint \ste{P}(t, t') e^{i\omega't'} \\
        &\quad \times e^{i(\omega - \omega')t} \,\text{d}t'\,\text{d}t.
        \end{split}
\end{equation}  
Substituting the expression of $\ste{P}(t,t-t')$ given by Eq.~(\ref{ptt}) in Eq.~(\ref{FTptt2}), and making use of the relation
\begin{equation} \label{orthog}
    \int e^{-i\om t}dt = 2\pi\delta(\om),
\end{equation}
where $\delta(\om)$ is the delta distribution, $\ste{P}(\om-\om',\om')$ can be found as
\begin{equation} \label{FTptt3}
    \ste{P}(\om-\om',\om') = \sum_n \ste{P}_n(\om')\delta(\om-n\Om-\om').
\end{equation}
We are also required to evaluate the Fourier transform of the field $\mathbb{F}(t)$. Under the Floquet ansatz, and noting that the original solutions are monochromatic with angular frequency $\bar{\om}$, the field can be expressed in phasor form as
\begin{equation} \label{F_tt}
    \mathbb{F}(t) = (2\pi)^{-1/2}\sum_n\text{Re}\left[\mathbb{F}_n (\bar{\om}) e^{-i\bar{\om} t}e^{-in\Om t}\right].
\end{equation}
We are interested to find Maxwell's equations for the $\mathbb{F}_n$ (the $\om$ dependence of  $\mathbb{F}_n$ will be omitted for brevity). Therefore, we work instead with the field
\begin{equation} \label{F_t}
    \mathbb{F}(t) = (2\pi)^{-1/2}\sum_n\mathbb{F}_n e^{-i\bar{\om} t}e^{-in\Om t}.
\end{equation}

Using Eq.~(\ref{FT}) and Eq.~(\ref{orthog}) together with Eq.~(\ref{F_t}) we obtain
\begin{equation} \label{F_w}
    \mathbb{F}(\om) = \sum_n\mathbb{F}_n \delta(\bar{\om}+n\Om-\om),
\end{equation}
i.e. the frequency spectrum of the field corresponds to an infinite set of sidebands of the frequency $\bar{\om}$. 
Upon inserting Eq.~(\ref{FTptt3}) in Eq.~(\ref{maxwell_compact}) [Eq.~(\ref{maxwell_compact_app})], the integral on the right-hand side becomes a summation, and we obtain
\begin{equation} \label{sum1}
    \frac{\om}{c}\sum_n \ste{P}_n(\om-n\Om)\mathbb{F}(\om-n\Om)- \ste{D}\mathbb{F}(\om)= 0.
\end{equation}
Upon substituting $\mathbb{F}(\om)$ from Eq.~(\ref{F_w}) in Eq.~(\ref{sum1}), we get
\begin{equation}
\begin{split}
    \frac{\om}{c}\sum_{n,n'} \ste{P}_n(\om_{-n})\mathbb{F}_{n'}\delta[\bar{\om}+(n+n')\Om-\om]&\\ -\ste{D}\sum_m\mathbb{F}_m \delta(\bar{\om}+m\Om-\om)= 0&.
\end{split}
\end{equation}

Since the sums run over all integers, we can make the replacement $n+n'=m$, which yields
\begin{equation} \label{sum2}
    \begin{split}
        \frac{\om}{c}\sum_{n,m} \ste{P}_n(\om-n\Om)\mathbb{F}_{m-n}\delta(\bar{\om}+m\Om-\om)&\\ -\ste{D}\sum_m\mathbb{F}_m \delta(\bar{\om}+m\Om-\om)= 0&.
    \end{split}
\end{equation}

Equation (\ref{sum2}) must be understood in the distributional sense. Integrating Eq.~(\ref{sum2}) over $\om$ gets rid of the delta distributions, giving 
\begin{equation}
   \sum_m \left[\frac{\om_m}{c}\sum_n \ste{P}_n(\om_{m-n})\mathbb{F}_{m-n}- \ste{D}\mathbb{F}_m\right]= 0,
\end{equation}
where we have made the dummy variable replacement $\bar{\om}\rightarrow \om$, and defined $\om_m \equiv \om+m\Om$. Lastly, since time harmonics are linearly independent, the equality above holds for each summand in the summation over $m$. Then, making the substitution $m-n\rightarrow n$ we finally arrive at Eq.~(\ref{maxwell_ptc}):
\begin{equation}
    \frac{\om_m}{c}\sum_n \ste{P}_{m-n}(\om_n)\mathbb{F}_{n}- \ste{D}\mathbb{F}_m= 0.
\end{equation}

\section{DERIVATION OF THE EIGENVALUE PROBLEM, EQ.~(\ref{RSE})} \label{AppD}
In this Appendix, we go through the steps required to arrive at our central result, Eq.~(\ref{RSE}). Our starting point is Eqs.~(\ref{green_current}-\ref{green_unmod}). Upon inserting Eq.~(\ref{green_unmod}) in Eq.~(\ref{green_current}), we obtain, via Eq.~(\ref{maxwell_inhom2}):
\begin{align} \label{rse1}
    \mathbb{F}_m(\mb{r}) &= -\frac{\om_m}{\om_m-\tilde{\om}_{\alpha}}\sum_{n\neq m,\alpha}\mathbb{F}_{\alpha}(\mb{r}) \nonumber \\
    &\quad \times \int \mathbb{F}_{\alpha}(\mb{r}')\cdot \ste{P}_{m-n}(\om_n)\cdot \mathbb{F}_n(\mb{r}')\text{d}\mb{r}'.
\end{align}
\raggedbottom
Therefore, the expansion of the modulated resonant states in terms of the static resonant states reads:
\begin{equation} \label{field_exp2}
    \mathbb{F}_m(\mb{r}) = \sum_{\alpha}c_{\alpha,m}\mathbb{F}_{\alpha}(\mb{r}),
\end{equation}
with the $c_{\alpha,m}$ given by 
\begin{equation} \label{rse2}
    c_{\alpha,m} = -\frac{\om_m}{\om_m-\tilde{\om}_{\alpha}}\int \mathbb{F}_{\alpha}(\mb{r}')\cdot \ste{P}_{m-n}(\om_n) \mathbb{F}_n(\mb{r}')\text{d}\mb{r}'.
\end{equation}
 Substituting Eq.~(\ref{field_exp2}) in Eq.~(\ref{rse2}), and making the replacement $\om\rightarrow\tilde{\om}$ to distinguish the eigenfrequencies from an arbitrary frequency, we arrive at the eigenvalue problem in Eq.~(\ref{RSE}):


\begin{equation} \label{rse3}
    (\tilde{\om}+m\Om-\tilde{\om}_{\alpha})c_{\alpha,m} = -(\tilde{\om}+m\Om)\sum_{\beta,n\neq m}V_{\alpha,\beta}^{n,m}c_{\beta,n},
\end{equation}

with $V_{\alpha,\beta}^{n,m}$ given by Eq.~(\ref{Vab}). 
\section{NUMERICAL CALCULATION OF THE RESONANT STATES OF THE TIME-MODULATED DIELECTRIC SLAB} \label{AppE}

The resonant states in the time-modulated dielectric slab can be found as the eigenfrequencies of its associated scattering matrix $\mathbb{S}(\om)$ \cite{globosits2024pseudounitary}, or alternatively, as zeros of its inverse, defining the eigenvalue problem \cite{tikhodeev2020influence,weiss2011strong,gippius2010resonant}:
\vspace{-0.03cm} 
\begin{equation} \label{smat1}
    \mathbb{S}^{-1}(\tilde{\om}_{\alpha}) \ket{O_{\alpha}} = 0,
\end{equation}
\vspace{-0.03cm} 
where $\ket{O_{\alpha}}$ are the (purely outgoing) amplitudes of the resonant state field outside the slab, for all harmonics \cite{zurita2010resonances,globosits2024pseudounitary}. $\mathbb{S}(\om)$ can be calculated with the harmonic balance method described in \cite{zurita2009reflection}. The solutions of Eq.~(\ref{smat1}) are then found using a custom Matlab code implementing the approach by Beyn and Bykov \cite{Beyn2012May,bykov2012numerical}, which can be used to solve the eigenvalue problem inside a pre-specified contour in the complex plane.  

\vspace*{\fill}  

\bibliography{bibliography_clean}

\end{document}